\documentclass[11pt]{article}
\usepackage{amsmath,amsfonts}

\newcommand{\ie}{i.e.,\ }
\newcommand{\eg}{e.g.,\ }

\newcommand{\ei}{\:\;}

\newcommand{\Dsl}{\ensuremath \:\raisebox{0.2ex}{\slash}\hspace{-0.74em} D}
\newcommand{\partialsl}{\ensuremath \,\raisebox{0.2ex}{\slash}\hspace{-0.58em} \partial}
\newcommand{\nsl}{\ensuremath \,\raisebox{0.2ex}{\slash}\hspace{-0.6em} n}

\newcommand{\nperd}{n\cdot \partial}

\newcommand{\tl}{\tilde{\lambda}}

\newcommand{\tD}{\tilde{D}}

\newcommand{\diag}{\operatorname{diag}}

\newcommand{\Lag}{\mathcal{L}}

\newcommand{\Lie}{\mathcal{L}}

\begin{document}
\begin{flushright}
NA-DSF-13/2008
\end{flushright}
\vspace{1ex}

\begin{center}
\textbf{\Large Very Special Relativity in Curved Space-Times}\\[4ex]%
Wolfgang M\"uck\\[3ex]%
\textit{Dipartimento di Scienze Fisiche, Universit\`a degli Studi di
  Napoli ``Federico II''\\ and INFN, Sezione di Napoli --- via Cintia, 80126 Napoli, Italy}\\%
E-mail: \texttt{mueck@na.infn.it}
\end{center}
\vspace{1ex}
\begin{abstract}
The generalization of Cohen and Glashow's Very Special Relativity to curved space-times is considered. Gauging the SIM(2) symmetry does not, in general, provide the coupling to the gravitational background. However, locally SIM(2) invariant Lagrangians can always be constructed. For space-times with SIM(2) holonomy, they describe chiral fermions propagating freely as massive particles.
\end{abstract}

\vspace{3ex}

\section{Introduction}
\label{intro}

Recently, Cohen and Glashow \cite{Cohen:2006ky,Cohen:2006ir} proposed an interesting origin of neutrino mass. Breaking Lorentz symmetry to a four-parameter subgroup called $SIM(2)$, the Dirac equation for a chiral fermion may be augmented with a non-local term leading to propagation as a massive particle. In this scheme, which they called Very Special Relativity (VSR), the departure from Lorentz-invariance implies the breaking of the discrete symmetries $P$, $CP$ and $T$ (but not $CPT$) suggesting a common origin for small $CP$-violating effects and neutrino masses. As Cohen and Glashow argued, VSR is consistent with current experimental bounds on Lorentz symmetry breaking and, therefore, constitutes an interesting modification of Standard Model physics.

In contrast to other approaches to Lorentz breaking, VSR is free of spurions, \ie it does not involve spontaneous symmetry breaking by non-zero expectation values of Lorentz tensors. This is a consequence of the fact that $SIM(2)$ does not possess invariant tensors (except the scalar), which implies also that a local Lagrangian, which breaks Lorentz symmetry while maintaining $SIM(2)$, cannot be constructed merely out of Lorentz tensors. Another feature of $SIM(2)$, which distinguishes it from its parent $SO(3,1)$, is that all its irreducible representations are one-dimensional, labelled by spin along a preferred axis. Hence, VSR predicts (very small) mass splittings in $SO(3,1)$ matter multiplets \cite{Fan:2006nd}. $SIM(2)$ supersymmetry has been considered in \cite{Cohen:2006sc,Lindstrom:2006xh}.

An obvious question is whether and in which ways VSR can be generalized to space-times, which are not Minkowski. Gibbons, Gomis and Pope \cite{Gibbons:2007iu} searched for deformations of $ISIM(2)$, \ie $SIM(2)$ plus translations, analogous in spirit to the deformation of the Poincar\'e to the de~Sitter or anti-de~Sitter algebras. They found that there is such a deformation, but space-time would be described by a Finslerian rather than Riemannian geometry. Alvarez and Vidal \cite{Alvarez:2008uy} considered implementations of $SIM(2)$ in de~Sitter space-time, motivated by the experimental finding of a small positive cosmological constant. To do this, they wrote down a local Lagrangian with auxiliary fields, which reproduces the VSR dynamics of the neutrino in Minkowski space-time, and replaced ordinary with covariant derivatives. They did not verify, however, whether the Lagrangian thus obtained is invariant under \emph{local} $SIM(2)$ transformations, as one would expect for a gauging. In the present paper, we intend to follow up on this point. As will be discussed, the gauging of $SIM(2)$ does not lead, in general, to a consistent coupling to gravity. However, in vacuum space-times with $SIM(2)$ holonomy, matters are similar to Minkowski space-time such that chiral neutrinos do propagate as massive particles.

An outline of the rest of the paper is as follows. In Sec.~\ref{sim2:flat}, the VSR dynamics of neutrinos in Minkowski space-time is reviewed, and $SIM(2)$ invariant Lagrangians leading to the VSR neutrino mass are given. In Sec.~\ref{sim2:general}, the gauging of $SIM(2)$ is discussed. Specializing to space-times with $SIM(2)$ holonomy, in where the $SIM(2)$ covariant derivate provides the complete coupling to the gravitational background, the propagation of a chiral fermion as a massive particle is derived. Finally, Sec.~\ref{sim2:conc} contains conclusions.

\section{VSR in Minkowski Space-Time}
\label{sim2:flat}

Consider the Lorentz algebra $SO(3,1)$ in the form 
\begin{equation}
 \label{sim2:Lorentz}
 [M_{ab},M_{cd}]= \eta_{ad}M_{bc} +\eta_{bc}M_{ad} -\eta_{ac}M_{bd} -\eta_{bd}M_{ac}~,
\end{equation}
with the metric $\eta_{ab} = \diag (+--\:-)$. In the vector representation, the generators are $(M_{ab})^c_{\ei d}=  2\delta_{[a}^c \eta_{b]d}$.
Throughout the paper, latin indices will denote the components in the local Lorentz frame, whereas greek indices label the space-time coordinates. In this section, the vierbein is fixed to $e^a_\mu=\delta^a_\mu$, but in the subsequent section this will be lifted.

In a light cone basis of $SO(3,1)$, defined by $M_{\pm i}= (M_{0i}\pm M_{3i})/\sqrt{2}$, $M_{-+}=M_{03}$, the generators of $SIM(2)$ are $J=M_{12}$, $K=M_{-+}$, $T_i=M_{+i}$ and satisfy the commutation relations
\begin{equation}
 \label{sim2:sim2alg}
  [T_i, T_j]=0~,\quad [J,T_i]=\epsilon_{ij}T_j~, \quad [K,T_i]= -T_i~,\quad [K,J]=0~.
\end{equation}
We shall write a general element of $SIM(2)$ in the compact form 
\begin{equation}
 \label{sim2:sim2}
  \frac12 \lambda^{ab} M_{ab} = \lambda J + \tl K +\lambda^i T_i~,
\end{equation}
where $\lambda=\lambda^{12}$, $\tl=\lambda^{-+}$, $\lambda^i=\lambda^{+i}$, remembering that $\lambda^{-i}=0$.
The essential property of $SIM(2)$, as defined above, is that it leaves invariant the \emph{direction} of the null vector $n^\mu$ with components $n^+=1$, $n^-=n^i=0$.
\begin{equation}
 \label{sim2:n.trafo}
  \delta n^\mu = \tl n^\mu~.
\end{equation}

The equation of motion for a chiral spinor containing the non-local $SIM(2)$ mass term is \cite{Cohen:2006ir}
\begin{equation}
 \label{sim2:eom.spinor}
  \left( \partialsl + \frac{m^2}2 \frac{\nsl}{\nperd} \right) \nu_L=0~.
\end{equation}
The field $\nu_L$ propagates as a particle of mass $m$, as one can easily see by squaring the left hand side of \eqref{sim2:eom.spinor} to a Klein-Gordon equation. The dynamics of \eqref{sim2:eom.spinor}
was obtained in \cite{Alvarez:2008uy} from a local Lagragian involving auxilliary fields, equivalent to the following,
\begin{equation}
 \label{sim2:Lag.flat}
  \Lag = \frac{i}{2} \bar{\nu}_L \partialsl \nu_L + i \bar{\chi}_L \nperd \psi_R
    +\frac12 m \left(\bar{\chi}_L\nsl \nu_L + \bar{\psi}_R \nu_L \right) + \text{c.c.}
\end{equation}
For the sake of clarity, we have indicated the handedness of the various spinor fields as subscripts. The fact that there is a local Lagrangian is not in contradiction to what was said in the introduction. Indeed, for obtaining $SIM(2)$ invariance of \eqref{sim2:Lag.flat} one must pay the price that $\chi_L$ is not a Lorentz spinor, as one can see from the \emph{global} $SIM(2)$ symmetries of \eqref{sim2:Lag.flat},
\begin{equation}
 \label{sim2:Lag.sym}
  \delta \nu_L = \frac14 \lambda^{ab} \gamma_{ab} \nu_L~, \quad
  \delta \psi_R = \frac14 \lambda^{ab} \gamma_{ab} \psi_R~, \quad
  \delta \chi_L = \frac14 \lambda^{ab} \gamma_{ab} \chi_L -\tl \chi_L~, \quad
  \delta n^\mu = \tl n^\mu~,
\end{equation}
where $\lambda^{ab}$ is defined by \eqref{sim2:sim2}. The transformations of the conjugate fields $\bar{\nu}$, $\bar{\psi}$ and $\bar{\chi}$ follow from \eqref{sim2:Lag.sym}.

There are other local Lagrangians involving auxiliary fields that give rise to the equation of motion \eqref{sim2:eom.spinor}. For example,
\begin{equation}
 \label{sim2:Lag1.flat}
  \Lag = \frac{i}{2} \bar{\nu}_L \partialsl \nu_L + i \bar{\chi}_L \nperd \nsl \chi_L
  + m \bar{\chi}_L \nsl \nu_L + \text{c.c.}
\end{equation}
As we have $\nperd \nsl= \frac12 \nsl \partialsl \nsl$, the auxiliary field $\chi_L$ appears in \eqref{sim2:Lag1.flat} only in the combination $\nsl \chi_L$. This suggests to consider $\nu_R=\nsl \chi_L$ and impose the constraint $\nsl \nu_R=0$ via a Lagrange multiplier, which is described by the Lagrangian
\begin{equation}
 \label{sim2:Lag2.flat}
  \Lag = \frac{i}{2} \bar{\nu} \partialsl \nu + \frac12 m \bar{\nu} \nu + \bar{\lambda}_R \nsl \nu_R + \text{c.c.}
\end{equation}
In this form of the Lagrangian, the neutrino field starts off as a Dirac spinor, $\nu=\nu_L+\nu_R$, with the usual Dirac mass term, but the right-handed component is constrained by the Lorentz breaking term. In the context of the Standard Model, one is naturally led to ask why $\nu_R$ should be sterile in the weak interactions. A similar question arises also in the other local Lagrangians, \eqref{sim2:Lag.flat} and \eqref{sim2:Lag1.flat}, where one may consider possible couplings of the auxiliary fields $\chi_L$ and $\psi_R$ to the weak interaction gauge fields. We shall not address these questions in this paper.

\section{SIM(2) in Curved Space-Times}
\label{sim2:general}

From this point on, let us consider general vierbeins $e^a_\mu$ with zero torsion. Our aim is to generalize the actions given in the previous section such that, first, they are manifestly coordinate invariant, and second,  the symmetry transformations \eqref{sim2:Lag.sym} are promoted to \emph{local} symmetries.

To start, let us make a little detour and review how space-time symmetries are treated in the tetrad formalism. This helps us to disentangle the space-time from the Lorentz-frame symmetries in Minkowski space-time and to obtain a local $SIM(2)$ frame symmetry.
A space-time symmetry is given by a Killing vector field $\xi^\mu$ satisfying
\begin{equation}
 \label{sim2:Killing}
  \Lie_\xi g_{\mu\nu} = \nabla_\mu \xi_\nu +\nabla_\nu \xi_\mu =0~.
\end{equation}
To promote this symmetry to a symmetry of the vierbeins, one combines the coordinate transformation $x'{}^\mu=x^\mu+\xi^\mu$ with a rotation of the local Lorentz frame, such that
\begin{equation}
 \label{sim2:Killing2}
  \delta_\xi e^a_\mu = -\Lie_\xi e^a_\mu +\lambda(\xi)^a_{\ei b} \, e^b_\mu 
  = -\left[ \xi^\nu \nabla_\nu e^a_\mu + (\nabla_\mu \xi^\nu) e^a_\nu \right] 
    +\lambda(\xi)^a_{\ei b} \,e^b_\mu =0~.
\end{equation}
Hence, one obtains
\begin{equation}
 \label{sim2:Lorentz.trafo}
  \lambda(\xi)_{ab} = -(\nabla_\mu \xi_\nu) e^\mu_a e^\nu_b -\xi^\mu \omega_{\mu ab}~,
\end{equation}
where $\omega_{\mu ab}$ are the spin connections determined by the zero torsion constraints $D_\mu e^a_\nu=\nabla_\mu e^a_\nu +\omega_\mu{}^a_{\ei b}\, e^b_\nu =0$. 

The presence of the null vector field $n^\mu$ (with \emph{frame} components $n^+=1$, $n^-=n^i=0$) breaks those space-time symmetries which lead to non-zero matrix elements $\lambda(\xi)^{-i}$ and $\lambda(\xi)^{-+}$. For example, for Minkowski space-time with a constant null vector, considered in the previous section, the remaining symmetries are $T_i$ and $J$, which generate an $E(2)$ subgroup of the Lorentz group, whereas $M_{-i}$ and $K$ are broken. This group can be enhanced to $SIM(2)$ containing $T_i$, $J$ and $K$ by \emph{postulating} that the null vector is always given by $n^\mu=e^\mu_+$. To achieve this, one modifies the transformation law of the frame components $n^a$ under $K$, giving rise to the $SIM(2)$ representation
\begin{equation}
 \label{sim2:n.rep}
 \Gamma^{(n)}(K)^a_{\ei b} = K^a_{\ei b} +\delta^a_b~,\quad 
 \Gamma^{(n)}(T_i)=T_i~, \quad \Gamma^{(n)}(J)=J~.
\end{equation}
It is straightforward to show that, in this representation, $n^a$ is $SIM(2)$ invariant. At this point, the $SO(3,1)$ of local Lorentz frame rotations has been reduced to $SIM(2)$, because the modified representation $\Gamma^{(n)}$ is not contained in a representation of $SO(3,1)$. It is this symmetry, which we would like to gauge.

In contrast to the usual tetrad formalism, where space-time tensors are invariant under frame rotations, $n^\mu$ transforms non-trivially under $SIM(2)$,
\begin{equation}
 \label{sim2:n.mu.trafo}
  \delta n^\mu = \delta e^\mu_+ = \lambda_+^{\ei a}\, e^\mu_a = \lambda^{-+}\, e^\mu_+ = \tl n^\mu~.
\end{equation}

Coupling $n^\mu$ to a Lorentz spinor $\nu$ makes it necessary to introduce an auxiliary field $\chi$ with an appropriate transformation law such that $(\bar{\nu} \nsl \chi)$ is $SIM(2)$-invariant. One finds, as in \eqref{sim2:Lag.sym},
\begin{equation}
 \label{sim2:chi.trafo}
  \delta \chi = \left( \frac14 \lambda^{ab} \gamma_{ab} -\tl \right) \chi~,
\end{equation}
implying the $SIM(2)$ representation 
\begin{equation}
 \label{sim2:chi.rep}
  \Gamma^{(\chi)}(K) = \frac12 \gamma_{-+} -1~,\quad 
  \Gamma^{(\chi)}(T_i)=\frac12 \gamma_{+i}~,\quad 
  \Gamma^{(\chi)}(J)=\frac12 \gamma_{12}~,
\end{equation}
where the terms formed by the gamma matrices are inherited from the spinor representation of $SO(3,1)$.

A $SIM(2)$ covariant derivative can be introduced as
\begin{equation}
 \label{sim2:cov.der}
  \tD_\mu = \nabla_\mu + \tilde{\omega}_\mu^{\ei +i}\,\Gamma(T_i) +\tilde{\omega}_\mu^{\ei -+}\,\Gamma(K) 
  + \tilde{\omega}_\mu^{\ei 12}\,\Gamma(J)~,
\end{equation}
where $\Gamma$ stands for the representation appropriate for the field the derivative acts on. We have adorned the derivative and the gauge fields with a tilde to distinguish them from the usual, $SO(3,1)$ covariant derivative and the spin connections, $D_\mu$ and $\omega_\mu^{\ei ab}$, respectively. Indeed, one cannot, in general, identify $\omega_\mu^{\ei ab}$ with $\tilde{\omega}_\mu^{\ei ab}$, because they do not form a closed set under $SIM(2)$ transformations. This can be seen, \eg in the transformation of the spin connection $\omega_\mu{}^{-+}$ under $SIM(2)$,
\begin{equation}
 \label{sim2:spin.conn.trafo}
  \delta \omega_\mu^{\ei -+} = - \partial_\mu \tl + \lambda^i\, \omega_{\mu}^{\ei -i}~.
\end{equation}
This involves $\omega_{\mu}^{\ei -i}$, for which there is no corresponding $SIM(2)$ transformation.

Hence, in general, the $SIM(2)$ gauge fields do not provide the coupling to gravity. However, for fields, which inherit their representation from $SO(3,1)$, we can use the $SO(3,1)$ covariant derivative, which does provide the coupling to gravity. In the case of the neutrino kinetic term, this is precisely what one wants to do, because the breaking of $SO(3,1)$ should come only from terms in the Lagrangian, which involve $n^\mu$. The unusual fields are $n^\mu$ and $\chi$, which transform in the representations \eqref{sim2:n.rep} and \eqref{sim2:chi.rep}, respectively, but one easily realizes that the combinations $\nsl \chi$ and $n^\mu \chi$ transform under $SIM(2)$ as $SO(3,1)$ fields. Hence, the $SO(3,1)$ covariant derivative may act on these combinations. Using the fields $\nu$, $\psi$ and $\chi$, transforming under $SIM(2)$ as in \eqref{sim2:Lag.sym}, one can write down the following, locally $SIM(2)$ invariant terms,
\begin{equation}
 \label{sim2:terms}
  \bar{\nu} \Dsl \nu~, \quad \bar{\psi}\Dsl \psi~, \quad 
  \bar{\chi} n^\mu D_\mu \psi~,\quad \bar{\chi} \nsl \Dsl \psi~,\quad 
  \bar{\chi} \nsl D_\mu n^\mu \chi~,\quad \bar{\chi} \nsl \Dsl \nsl \chi~,\quad
  \bar{\chi} \nsl \nu~,\quad \bar{\psi} \nu~,
\end{equation}
as well as their complex conjugates. Other $SIM(2)$ invariant terms, for example $\bar{\chi} \gamma^{\mu\nu\rho} n_\mu (\nabla_\nu n_\rho) \chi$, are equivalent to these. As an aside, we remark that the Lagrangian given in eq.~(12) of \cite{Alvarez:2008uy} is not locally $SIM(2)$ invariant, because the third term on the right hand side of that equation is not.

In what follows, we shall assume that $\omega_\mu^{\ei -i}=0$, in which case we can identify the $SIM(2)$ gauge fields with the remaining spin connections, and $D_\mu$ agrees with $\tD_\mu$ when acting on Lorentz fields,
\begin{equation}
 \label{sim2:omega.-i}
  \omega_\mu^{\ei -i}=0~:\qquad \tilde{\omega}_\mu^{\ei ab}=\omega_\mu^{\ei ab}~,\qquad D_\mu \to \tD_\mu~.
\end{equation}
We should interpret this assumption in the sense that we consider those space-times, in which one can choose a Lorentz frame such that \eqref{sim2:omega.-i} holds. This is not the generic case, since the six $SO(3,1)$ frame rotations do not suffice to eliminate the eight components $\omega_\mu^{\ei -i}$. If, however, such a frame exists, then the condition \eqref{sim2:omega.-i} is $SIM(2)$ invariant.

The assumption \eqref{sim2:omega.-i} can be rephrased as $SIM(2)$ holonomy. This follows from the fact that the $SIM(2)$ covariant derivative of $n^\mu$ vanishes implying that $n^\mu$ is a recurrent null vector field,
\begin{equation}
 \label{sim2:tDn}
  (\tD_\mu n^\nu) = (\nabla_\mu n^\nu) + \omega_\mu^{\ei-+}\, n^\nu=0~.
\end{equation}
A corollary of \eqref{sim2:omega.-i} is
\begin{equation}
 \label{sim2:R.-i}
  R_{\mu\nu}^{\ei\ei -i} = 0~.
\end{equation}
For more information on space-times with $SIM(2)$ holonomy, we refer to \cite{Gibbons:2007zu} and references therein. 

In the remainder, we shall consider the generalization of VSR to space-times with $SIM(2)$ holonomy, \ie satisfying \eqref{sim2:omega.-i}. In addition, we assume the space-time to be a vacuum solution of Einstein's equations, possibly with a cosmological constant, such that $R_{\mu\nu}=\Lambda g_{\mu\nu}$.
The commutator of two $SIM(2)$ covariant derivatives, which will be used below, reflects the $SIM(2)$ holonomy,
\begin{equation}
  \label{sim2:tDD}
  \left[\tD_\mu, \tD_\nu \right] = \left[\nabla_\mu, \nabla_\nu\right] + R_{\mu\nu}^{\ei\ei -+}\, \Gamma(K)
    + R_{\mu\nu}^{\ei\ei 12}\, \Gamma(J) + R_{\mu\nu}^{\ei\ei +i}\, \Gamma(T_i)~.
\end{equation}

Let us consider the simplest Lagrangian, which is given by \eqref{sim2:Lag2.flat}, with $\partial_\mu$ replaced by $D_\mu=\tD_\mu$ (as it acts on a Lorentz spinor),
 \begin{equation}
 \label{sim2:Lag.gen}
  \Lag = \frac{i}{2} \bar{\nu} \tilde{\Dsl} \nu + \frac12 m \bar{\nu} \nu + \bar{\lambda}_R \nsl \nu_R + \text{c.c.}
\end{equation}
It gives rise to the equations of motion
\begin{align}
 \label{sim2:eom.a}
  i \tilde{\Dsl} \nu_L + m \nu_R &=0~,\\
 \label{sim2:eom.b}
  i \tilde{\Dsl} \nu_R + m \nu_L +\nsl \lambda_R &=0~,\\
 \label{sim2:eom.c}
  \nsl \nu_R &=0~.
\end{align}
After multiplying \eqref{sim2:eom.b} by $\nsl$ and using \eqref{sim2:tDn} and \eqref{sim2:eom.c}, one obtains
\begin{equation}
 \label{sim2:nuR.sol}
  \nu_R= \frac{im}{2n^\mu \tD_\mu} \nsl \nu_L~.
\end{equation}
Substituting \eqref{sim2:nuR.sol} back into \eqref{sim2:eom.b} and making use of $\nsl \tilde{\Dsl} \nu_L=0$, which follows from \eqref{sim2:eom.a}, yields
\begin{equation}
 \label{sim2:n.lambda}
  \nsl \lambda_R = \frac{i}{n^\rho \tD_\rho} \gamma^\mu n^\nu \left[\tD_\mu, \tD_\nu \right] \nu_R~.
\end{equation}
As $\nu_R$ is a Lorentz spinor and satisfies $\gamma_+\nu_R=\nsl \nu_R=0$, \eqref{sim2:n.lambda} becomes
\begin{equation}
 \label{sim2:n.lambda2}
  \nsl \lambda_R = \frac{i}{2n^\rho \tD_\rho} \gamma^\mu n^\nu 
    \left( R_{\mu\nu}^{\ei \ei 12}\, \gamma_{12} -R_{\mu\nu}^{\ei \ei -+} \right) \nu_R
  = \frac{i}{2n^\mu \tD_\mu} R_{+j} \, \gamma^j \nu_R =0~.
\end{equation}
The last step follows from the vacuum property of the space-time background.
Finally, from \eqref{sim2:eom.a}, \eqref{sim2:eom.b} and \eqref{sim2:n.lambda2} easily follows
\begin{equation}
 \label{sim2:propa.gen}
  \left( \Dsl\Dsl + m^2 \right) \nu_L = 0~.
\end{equation}

\section{Conclusions}
\label{sim2:conc}

In this paper, the generalization of Cohen and Glashow's Very Special Relativity to curved space-times has been considered. In general, gauging the $SIM(2)$ symmetry, which leaves the preferred null direction $n^\mu$ invariant, does not provide the complete couplings to the gravitational background. One can, however, construct locally $SIM(2)$ invariant Lagrangians from the terms listed in \eqref{sim2:terms}. These terms make use of the standard $SO(3,1)$ covariant derivative and, therefore, do not derive from a standard gauging of $SIM(2)$. Moreover, for a general space-time and/or a generic null vector field $n^\mu$, such Lagrangians do not lead to freely propagating chiral fermions. Instead, for space-times with $SIM(2)$ holonomy, the $SO(3,1)$ covariant derivatives in the Lagrangians coincide with $SIM(2)$ covariant derivatives. In these cases, and if, in addition, the space-time is a vacuum, the Lagrangians describe freely propagating massive chiral fermions, just as in Minkowski space-time. It is essential here that the null vector field $n^\mu$ is not generic, but is such that its direction remains invariant under parallel transport.

\section*{Acknowledgments}
It is a pleasure to thank L.~Cappiello for fruitful discussions.
This work has been supported in part by the European Community's Human Potential Programme under
contract MRTN-CT-2004-005104 'Constituents, fundamental forces and symmetries of the 
universe' and by the Italian Ministry of Education and Research (MIUR), project 2005-023102.

\bibliographystyle{JHEP}
\bibliography{sim2}

\end{document}